\documentclass[letterpaper, 10 pt, conference]{ieeeconf}

\IEEEoverridecommandlockouts                              

\usepackage[backend=biber, sorting=none, style=numeric-comp]{biblatex}
\usepackage{xparse}
\renewcommand{\baselinestretch}{0.98} 

\let\labelindent\relax
\usepackage[inline]{enumitem}

\usepackage{tcolorbox} 

\usepackage{graphics} 
\usepackage{epsfig} 
\usepackage{times} 
\usepackage{amsmath} 
\allowdisplaybreaks
\usepackage{amssymb}  

\usepackage{bm}

\usepackage{amsfonts}
\usepackage{wasysym}
\usepackage{pifont}	

\usepackage[ruled,vlined]{algorithm2e}

\usepackage{booktabs}
\usepackage{makecell}

\renewcommand{\abovecaptionskip}{0pt}

\usepackage[hyperindex=true, %
  pdftitle={Real-time Projected Gradient-based Nonlinear Model Predictive Control with an Application to Anesthesia Control},%
  pdfauthor={Sophie Hall, Lukas Ortmann, Miguel Picallo, Florian D\"{o}rfler},%
  colorlinks=true,%
  pagebackref=false,%
  plainpages=false,%
  pdfpagelabels,%
  linkcolor=black,%
  citecolor=black,%
  filecolor=black,%
  urlcolor=black%
  ]{hyperref}

\definecolor{lightgray}{gray}{0.9}
\definecolor{ggreen}{rgb}{0,0.5,0}
\definecolor{rred}{rgb}{0.8,0,0} 
\definecolor{bblue}{rgb}{0.01, 0.31, 0.59}

\newcommand{\bb}[1]{\mathbb{#1}}

\newcommand{\re}[1]{{\color{black}{#1}}}  

\DeclareMathOperator*{\argmin}{arg min} 

\newtheorem{assump}{Assumption}
\newtheorem{theorem}{Theorem}
\newtheorem{remark}{Remark}
\newtheorem{prop}{Proposition}

\newtheorem{cor}{Corollary}
\newtheorem{lemma}{Lemma}

\graphicspath{{graphics/}}
\title{\LARGE \bf
Real-time Projected Gradient-based Nonlinear Model Predictive Control with an Application to Anesthesia Control}

\author{Sophie Hall, Lukas Ortmann, Miguel Picallo, Florian D\"{o}rfler
\thanks{The authors are with the Automatic Control Laboratory, ETH Z\"urich, Physikstrasse 3, 8092 Z\"urich, Switzerland. This research has been supported by ETH Zürich funds and the NCCR Automation.}%
\thanks{Email: \, {\tt\footnotesize \{shall,ortmannl,miguelp,doerfler\}@ethz.ch}}}

\begin{document}

\maketitle
\thispagestyle{empty}
\pagestyle{empty}

\begin{abstract}

\re{Medical drug infusion problems} pose a combination of challenges such as nonlinearities from physiological models, model uncertainty due to inter- and intra-patient variability, as well as strict safety specifications. With these challenges in mind, we propose a novel real-time Nonlinear Model Predictive Control (NMPC) scheme based on projected gradient descent iterations. At each iteration, a small number of steps along the gradient of the NMPC cost is taken, generating a suboptimal input which asymptotically converges to the optimal input. We retrieve classical Lyapunov stability guarantees by performing a sufficient number of gradient iterations until fulfilling a stopping criteria. \re{Such a real-time control approach allows for higher sampling rates and faster feedback from the system which is advantageous for the class of highly variable and uncertain drug infusion problems.} To demonstrate the controller's potential, we apply it to hypnosis control in anesthesia of two interacting drugs. The controller successfully regulates hypnosis even under disturbances and uncertainty and fulfils benchmark performance criteria.

\end{abstract}

\section{INTRODUCTION}

Health care costs are rising in many countries and an ageing population requires more frequent and complex treatments \cite{papanicolas2018health}. 
Medical control systems have the potential to ensure adequate and timely therapy delivery while allowing clinicians to focus on fewer more complex tasks \cite{parvinian2018regulatory}. However, medical control poses various challenges such as \begin{enumerate*}
\item nonlinearities from physiological models, \item model uncertainty due to inter- and intra-patient variability, and \item strict safety specifications \cite{ goodwin2019applications} \end{enumerate*}. 
In this paper we specifically focus on the class of automated drug infusion problems which have received considerable attention in recent times as they can be modelled well \cite{macheras2006modeling} and cover relevant applications such as control of:
Blood pressure \cite{yu2006blood}, hemoglobin \cite{gaweda2008model}, haemodialysis \cite{javed2011recent}, anesthesia~\cite{ntouskas2021robust}, and others~\cite{khodaei2019physiological}.

A common control approach for medical problems is Model Predictive Control (MPC) which uses dynamic models to predict and optimize system behaviour \cite{goodwin2019applications}. MPC offers flexibility in formulating the objective and allows for systematic incorporation of constraints \cite{rawlings2017model}. Nonlinear MPC (NMPC) can control nonlinear dynamics making it a good approach for medical control problems \cite{goodwin2019applications}.

However, the implementation of MPC requires solving a constrained (and potentially non-convex) optimal control problem (OCP) at each time step which may be prohibitively computationally expensive \cite{rawlings2017model, gruene2017nonlinear} and may lead to slow sampling rates. \re{Especially in embedded systems used for drug infusion problems, such as an artificial pancreas, computational resources are limited \cite{zavitsanou2016embedded}. Furthermore, in applications with high model uncertainty due to inter- and intra- patient variability, fast feedback from the system is essential to mitigate set-point deviations. In order to reduce computational complexity,  approaches such as explicit MPC and real-time MPC have been proposed.}

In explicit MPC the feedback law is precomputed offline and stored as a lookup table. It has been widely adopted in the medical field \cite{ingole2017offset, nascu2015offset, krieger2013modelling}. However, explicit MPC does not scale well with increasing dimensionality caused by long prediction horizons required in drug infusion problems. In real-time MPC first- and second-order optimization algorithms are employed to approximate the solution of an MPC problem. At each time-step a certain number optimization steps are taken which improve upon an initial solution guess, iteratively converging to the optimal solution. Nevertheless, analysing stability under a suboptimal input is challenging. Yet stability certificates are especially important in safety-critical applications such as medical control. There is an extensive literature on real-time MPC~\cite{diehl2005nominal,graichen2010stability, allan2017inherent}. An overview of different methods is given in~\cite{gruene2017nonlinear} and~\cite{liaomcpherson2020time}.

\re{Not only, does real-time MPC reduce computational complexity but additionally in applications with large model uncertainty, it has been shown that the optimal solution is very sensitive to parameter changes and applying first-order methods can be more robust  \cite{doerfler2019distributed}. Thus, real-time MPC is an ideal candidate for solving automated drug infusion problems as it allows for fast feedback from the system and increases robustness, both important properties for applications with high inter- and intra patient variability.}


Consequently, in this work we present a novel real-time projected gradient-based NMPC scheme particularly suited for drug infusion problems. 
Our contributions are threefold: First, with the novel real-time NMPC scheme we successfully control hypnosis in anesthesia of two interacting medications fulfilling predefined performance criteria. To the best of our knowledge this is the first work applying real-time NMPC to a drug infusion problem. 
Second, our theoretical analysis of the proposed scheme allows us to show stability under standard NMPC assumptions. We extend existing theory for real-time NMPC by deriving a theoretical stopping criterion rather than fixing the number of iterations of the optimization algorithm which is done in previous works \cite{diehl2005nominal, liaomcpherson2020time, graichen2010stability}. Such a stopping criterion allows to perform just enough iterations to preserve stability, e.g., only few iterations when there are no significant state changes, but many iterations in the presence of disturbances which would destabilise the system otherwise. Third, we present a self-contained tutorial on modelling of anesthesia control.
 

The paper is organized as follows. We give preliminaries in Section~\ref{sct:Preliminaries} and present the projected gradient-based controller in Section~\ref{sct:Controller}. Simulation results for anesthesia control are given in Section~\ref{sct:Anesthesia}. Section \ref{sct:Conclusions} concludes the paper. 

\section{PRELIMINARIES ON NMPC} \label{sct:Preliminaries}

Consider a nonlinear discrete-time system
\begin{equation} 
x_{t+1}= f(x_t,u_t) , 
\label{dfn:Dynamics}
\end{equation} 

\noindent
where $x_t \in \mathbb{R}^{n_x}$ represents the system state at $t$ and $u_t \in \mathcal{U} \in \mathbb{R}^{n_u}$ is the control input. We assume the function $f: \bb{R}^{n_x}\times \mathbb{R}^{n_u} \to \bb{R}^{n_x}$ is continuously differentiable and that without loss of generality the origin is an equilibrium $f(0,0)= 0$. If not stated otherwise, $\|\cdot\|$ denotes the Euclidean norm. We consider a typical NMPC setup in which we use a nonlinear model of the system to predict and influence future system behaviour over a time horizon $i=1,\dots,N$. We want to minimize the running cost $h:\bb{R}^{n_x}\times \bb{R}^{n_u\times N}\to \bb{R} $ which is comprised of a stage cost $l: \bb{R}^{n_x}\times \bb{R}^{n_u}\to \bb{R}$ and a terminal cost $V_f: \bb{R}^{n_x}\to \bb{R}$. We define $x$ to be the most recent measured state. Further, we define the predicted state and future input sequence as $\bm{\xi} = [\xi_0,\dots, \xi_{N}]$ and $\bm{\mu} = [\mu_0,\dots, \mu_{N-1}]$.  Only the inputs $\mu_k$ are optimization variables and not the predicted states $\xi_k$ which are a function of $x$ and $[\mu_0,\dots, \mu_{k-1}]$: $\xi_k(x, \mu_0, \dots, \mu_{k-1}) = f(\cdot, \mu_{k-1}) \circ \dots \circ f(\cdot, \mu_{1})\circ f(x, \mu_{0}), \; \forall k\geq 1,$ and $\xi_0 = x$.  The running cost $h(x,\bm{\mu})$ and the optimal control sequence $\bm{\mu}^*(x)$ are defined as

\vspace{-0.7cm}
\begin{equation}
\begin{array}{lrrlll}
\bm{\mu}^*(x)= &\displaystyle\argmin_{\bm{\mu}} & \multicolumn{4}{l}{\overbrace{\sum_{k=0}^{N-1} l(\xi_k, \mu_k)+ V_f(\xi_N)}^{h(x,\bm{\mu})}} \\

&\textrm{s.t.} & \mu_k  \in \mathcal{U} &\; k= 0, ..., N-1,
\\
\end{array}\label{dfn:OP}
\end{equation}

\noindent and the optimal control input is the first element of the sequence $\bm{\mu}^*(x)$, i.e., $u^*(x)= \mu^*_0(x)$. The optimal value function $V(x) = h(x, \bm{\mu}^*(x))$ gives the optimal value of (\ref{dfn:OP}).


We assume the following to ensure that the NMPC problem is well-posed, similar to \cite{diehl2005nominal} 
and \cite{diehl2007stabilizing}: 
\vspace{3pt}
\begin{assump}\label{assump:NMPC1}
There exists an open set $\mathcal{X}\subset \mathbb{R}^{n_x}$ such that $\forall x \in \mathcal{X}$, the OCP (\ref{dfn:OP}) is feasible and has a unique optimal solution $\bm{\mu}^*(x)$. Further, $V(x)$ is continuous and upper-bounded: $V(x)<M\|x\|^2, \, \forall x\in \mathcal{X} \backslash \{0\}$ with $M>0$.
\end{assump}

\vspace{0.1cm}
We further define the set $\mathcal{X}_{\alpha}$ which is for a fixed $\alpha>0$ the maximum sub-level set of $V(x)$ contained in $\mathcal{X}$,
\begin{equation}
\mathcal{X}_{\alpha} := \{ x \in \mathcal{X}\,|\, V(x)\leq \alpha \} \subset \mathcal{X}.
\label{dfn:Xalpha}
\end{equation}

We make the following common MPC assumption on the running cost $h(x,\bm{\mu})$ and the constraints \cite[Ass. 2.2, 2.3]{rawlings2017model}:

\vspace{0.1cm}
\begin{assump} \label{assump:NMPC2}
The stage cost and terminal cost satisfy $l(x,u)>0,  \; \forall (x, u) \in \mathcal{X} \times \mathcal{U}\backslash \{(0,0)\}$ and $ V_f(x)>0$, $ \forall x \in~\mathcal{X}\backslash \{0\}$ as well as $l(0,0) = 0$ and $V_f(0)=0$. Also, there exists $d>0$ such that $l(x,u)\geq d\,, \forall x \notin \mathcal{X}_{\alpha}$. Further, the input constraint set $\mathcal{U}$ is closed and convex. 
\end{assump}

In nominal NMPC the full optimization problem (\ref{dfn:OP}) is solved in receding horizon and to guarantee stability the following standard NMPC assumption is required~\cite{diehl2007stabilizing}. 

\vspace{0.1cm}

\begin{assump} \label{assump:NMPC3}  There exists a local controller $\kappa(x) \in~ \mathcal{U}, \; \forall x \in \mathcal{X}_{\alpha}$ such that $f(x, \kappa(x)) \in \mathcal{X}_{\alpha}, \forall x \in \mathcal{X}_{\alpha}$ and 
 $V_f(f(x, \kappa(x)))- V_f(x) \leq -l(x, \kappa(x)), \forall x \in \mathcal{X}_{\alpha}  \backslash \{0\}$.
\end{assump}
\vspace{0.1cm}

This allows us to state the following well-known Theorem:

\begin{theorem}\cite[Theorem 2.19]{rawlings2017model} \label{thrm:NomStab}Under Assumptions~\ref{assump:NMPC1}-\ref{assump:NMPC3} it holds $\forall x \in \mathcal{X}$:
 \begin{align} \label{eqn:NMPCDecrease}
&V(f(x,u^*(x)))\leq V(x)- l(x, u^*(x)),
\end{align}
and the origin is asymptotically stable in $\mathcal{X}$ for $f(x, \kappa(x))$.
\end{theorem}

A common approach to ensure recursive feasibility in NMPC is to add a terminal constraint such that $\xi_{N} $ lies in $\mathcal{X}_{\alpha}$ in which the terminal controller $\kappa(\xi_{N})$ is defined. However, the OCP (\ref{dfn:OP}) doesn't enforce state constraints, thus to ensure recursive feasibility we instead make use of the following Theorem characterizing the domain of attraction without terminal constraint.


\vspace{0.1cm}
\begin{theorem}\cite[Theorem 1]{limon2006stability}  \label{thrm:Gamma}
Under Assumptions~\ref{assump:NMPC1}-\ref{assump:NMPC3}, there exists a set $\Gamma$, in which the optimal controller (\ref{dfn:OP}) satisfies the cost decrease (\ref{eqn:NMPCDecrease}) and asymptotically stabilizes (\ref{dfn:Dynamics}) for all $ x \in \Gamma$. The set $\Gamma$ is defined as
\begin{equation} \label{dfn:Gamma}
\Gamma = \{x \in \mathbb{R}^{n_x}: V(x)\leq N  \cdot d + \alpha\} \subset \mathcal{X},
\end{equation}
where $N\in \mathbb{N}$ is the horizon length, $d$ is the lower bound given in Assumption \ref{assump:NMPC2}, and $\alpha$ is defined in (\ref{dfn:Xalpha}).
\end{theorem}

As $V(x)$ is positive and continuous on $x$ and the interval $[0, N\cdot d + \alpha]$ is closed, the set $\Gamma \subset \bb{R}^n$ is closed, bounded and thus compact. Note that the following holds: $\mathcal{X}_{\alpha} \subset \Gamma \subset \mathcal{X}$.

\section{REAL-TIME PROJECTED GRADIENT-BASED NMPC} \label{sct:Controller}

NMPC schemes give rise to complex optimization problems like (\ref{dfn:OP}). One approach to reduce computation is to employ a real-time iteration scheme. Rather than solving the OCP (\ref{dfn:OP}) at every time step $t$, an estimate of the optimal solution is maintained and improved upon by applying a finite number of iterations of an optimization algorithm, like in~\cite{diehl2005nominal,diehl2007stabilizing, liaomcpherson2020time}. 

In this paper we present an approach in which a finite number of projected gradient steps are applied to a shifted version of the input sequence $\bm{\mu}_{t-1}=[\mu_{0|t-1}, \dots, \mu_{N-1|t-1}]$ where the element $\mu_{k|t-1}$ is the input for $k$ steps into the future computed at time $t-1$. The first element of the newly generated sequence $\bm{\mu}_{t}$ is applied to the system. More specifically this procedure is summarized in Algorithm~\ref{algo:Algorithm1}.

\vspace{-0.25cm}
\begin{algorithm}[h]
\LinesNumbered
\KwIn{$x_t, x_{t-1}, \bm{\mu}_{t-1}, \gamma, \epsilon, \sigma$}
\textbf{Initialize} $ x = x_t, \newline \bm{\mu} = [\mu_{1|t-1}, \dots, \mu_{N-1|t-1}, \kappa(\xi_N(x_{t-1}, \bm{\mu}_{t-1}))],$\\ 
 \While{$ \| \bm{\mu}- \Pi_{\mathcal{U}^N}\left[ \bm{\mu} - \gamma \nabla_{\bm{\mu}} h(x,\bm{\mu})\right]\|\geq \frac{\sqrt{1- \epsilon^2}}{\sigma}  \;l(x,\mu_0) $}{
 $\; \bm{\mu} =  \Pi_{\mathcal{U}^N}(\bm{\mu} - \gamma \nabla_{\bm{\mu}} h(x,\bm{\mu})) $ }
\KwOut{$\bm{\mu}_{t} = \bm{\mu}$}
Apply $u_t = \mu_0$ to system\\
 \caption{\strut Real-time projected gradient NMPC} \label{algo:Algorithm1}
\end{algorithm}
\vspace{-0.3cm}

In Algorithm~\ref{algo:Algorithm1} $\gamma$ is the fixed step size, and the Euclidean projection is $\Pi_{\mathcal{U}^N}(\bm{\mu}):= \argmin_{y \in \mathcal{U}^N} \|\bm{\mu}-y\|^2$ with $\mathcal{U}^N = \mathcal{U} \times ... \times \mathcal{U}$. During the initialisation of Algorithm~\ref{algo:Algorithm1} we make use of a common warm-starting technique in NMPC~\cite{gruene2017nonlinear}: A new input sequence is generated by shifting the previous sequence and adding a local controller at the end. The stopping criterion is required to ensure stability of the real-time NMPC scheme in Algorithm~\ref{algo:Algorithm1} and it is given by $StopCrit(x, \bm{\mu}) = \frac{\sqrt{1- \epsilon^2}}{\sigma} \, l(x,u)$\re{, where $\epsilon$ and $\sigma$ are constants which depend on the contraction rate and Lipschitz constants of the system.} The detailed derivation is given in Section~\ref{scn:Proof}. Additionally for stability we require the following assumption.
 
\vspace{0.05cm}
\re{\begin{assump}\label{assump:RunningCost}
The set $\Gamma$ resulting from Theorem 2 under Assumption 1- 3 satisfies: $l(x,u)$ is $L_1$ Lipschitz continuous $\forall (x,u) \in \Gamma \times \mathcal{U}$, $V_f(x)$ is Lipschitz continuous $\forall x \in \Gamma$ and $h(x,\bm{\mu})$ is $m$-strongly convex $\forall x \in \Gamma$.
\end{assump}}

\begin{remark}
Assumption~\ref{assump:RunningCost} is similar to assumptions made in related works such as \cite{liaomcpherson2020time}. If $h(x,\bm{\mu})$ was not convex in $\Gamma$, a smaller sublevel set of $V(x)$ may be considered and the same analysis still holds.
\end{remark}
Assumption~\ref{assump:RunningCost} ensures contractivity of the real-time iterations under the suboptimal input $u$ in Algorithm~\ref{algo:Algorithm1}.  We can conclude the following Lemma:

\begin{lemma}
Under Assumption~\ref{assump:RunningCost} the running cost $h(x, \bm{\mu})$ is $L_2$ Lipschitz continuous $\forall (x,\bm{\mu}) \in \Gamma \times \mathcal{U}^N$.
\end{lemma}

\textit{Proof:} The running cost $h(x, \bm{\mu})$ is a composition of Lipschitz functions $l(x,u)$, $V_f(x,u)$ and $f(x,u)$, resulting in $h(x, \bm{\mu})$ being Lipschitz \cite[Proposition 2.3.1]{cobzas2019lipschitz}.

\vspace{0.15cm}
Additionally, we require the dynamics and the optimal solution to fulfil the following assumption: 

\begin{assump} 
The local solution $\bm{\mu}^*(x)$ in (\ref{dfn:OP}) fulfils $\forall x \in \Gamma$ \begin{enumerate*}[label=(\roman*)]
  \item the linear independence condition and 
  \item the second order sufficiency condition 
\end{enumerate*} \cite{jittorntrum1978sequential}.
\label{assump:Estmu}
\end{assump}

Assumption~\ref{assump:Estmu} allows us to derive a Lipschitz bound between $\bm{\mu}^*(f(x,u))$ and $\bm{\mu}^*(f(x,u^*(x))$ with the corresponding Lipschitz constant $L_3$. 

To use the value function $V(x)$ as a Lyapunov function, a decrease at every time step $t$ under the suboptimal input $u$ is required which is established by the following Theorem:

\vspace{0.15cm}

\re{
\begin{theorem} \label{thrm:ValueFuncDecrease}
Under Assumptions~\ref{assump:RunningCost}-\ref{assump:Estmu}, if it holds that $ \| \bm{\mu}- \Pi_{\mathcal{U}^N}\left[ \bm{\mu} - \gamma \nabla_{\bm{\mu}} h(x,\bm{\mu})\right]\|\leq \frac{\sqrt{1- \epsilon^2}}{\sigma} \, l(x,u) $, then $\forall x~\in~\Gamma\backslash \{0\}$ $V(f(x,u))< V(x)$.
\end{theorem} }
\vspace{0.15cm}

The definition of $\Gamma$ is given in (\ref{dfn:Gamma}) and the detailed derivation of Theorem~\ref{thrm:ValueFuncDecrease} is given in the upcoming section. Theorem~\ref{thrm:ValueFuncDecrease} allows to conclude the following Corollary:
 
\vspace{0.15cm}
\re{\begin{cor}
Under Assumptions~\ref{assump:RunningCost}-\ref{assump:Estmu}, $\forall x\in~\Gamma$ the system~(\ref{dfn:Dynamics}) under Algorithm~\ref{algo:Algorithm1} is asymptotically stable.
\end{cor} }

\vspace{0.15cm}

In general, for asymptotic stability the number of gradient steps $-\gamma \nabla_{\bm{\mu}} h(x,\bm{\mu})$ taken must be large enough to fulfil the stopping criterion $StopCrit(x, \bm{\mu})$. However, depending on the application one might also choose to perform more gradient updates for faster convergence to the optimal input sequence $\bm{\mu}^*(x)$. This of course comes at a higher computational burden per iteration $t$. Overall, an advantage of our real-time projected gradient-based NMPC scheme is that it gives a direct handle for balancing computational complexity and performance while ensuring stability.
\subsection{Proof of Theorem \ref{thrm:ValueFuncDecrease}} \label{scn:Proof}

The stability of Algorithm~\ref{algo:Algorithm1} is shown through two main steps. Firstly, we show that with every iteration of the projected gradient algorithm the distance between the optimal input sequence $\bm{\mu}^*(x)$ and the suboptimal one $\bm{\mu}$ decreases. Secondly, we show that standard NMPC Lyapunov stability theory is still applicable if $u^*(x)$ and $u$ are close enough which we assure with the stopping criteria.
\vspace{0.1cm}


\subsubsection{Contractivity of real-time iterations}

Given Assumption~\ref{assump:RunningCost}, we can derive relations between the suboptimal inputs in Algorithm~\ref{algo:Algorithm1} step 3 and the optimal input $\bm{\mu}^*(x)$.

\vspace{0.15cm}

\begin{theorem}\cite[Theorem 1]{vanparys2019real} \label{thm:RealTimeIt} Under Assumption~\ref{assump:RunningCost}, for a single gradient update $\bm{\mu}^{i+1} = \Pi_{\mathcal{U}^N}[\bm{\mu}^i-\gamma \nabla_{\bm{\mu}} h(x, \bm{\mu}^i)]$ with $\gamma \in (0, 2/L_2)$ and $\epsilon^2= (1-\frac{2m^2}{L_2}\gamma + m^2\gamma^2)$ it holds:
\vspace{0.1cm}
\begin{itemize}
\item Decreasing distance to local optimizer
\begin{equation}
\|\bm{\mu}^{i+1}-\bm{\mu}^*\|^2 \leq \epsilon^2 \|\bm{\mu}^i - \bm{\mu}^{*}\|^2 
\end{equation}

\item Bounding distance to local optimizer
\begin{equation} \label{eqn:DistOPt}
(1-\epsilon^2) \|\bm{\mu}^i-\bm{\mu}^*\|^2 \leq \|\bm{\mu}^i-\bm{\mu}^{i+1}\|^2
\end{equation}


\end{itemize}

\end{theorem}

As we require $\gamma \in (0, 2/L_2)$, it follows $\epsilon^2 \in (0,1)$. Theorem~\ref{thm:RealTimeIt} shows that the distance to the optimal input sequence $\bm{\mu}^*(x)$ decreases which is required to ensure a decrease in the optimal cost $V(x)$.

\vspace{0.15cm}
\subsubsection{Decrease in optimal cost $V(x)$}
As shown in (\ref{eqn:NMPCDecrease}), the cost of an optimal NMPC decreases by at least $-l(x, u^*(x))$ at every time step.  However, in our real-time projected gradient-based scheme we do not apply the optimal input $u^*(x)$ to the system. Consequently, the question is if a decrease in the optimal value function $V(x)$ is ensured even when applying the suboptimal input $u$. The following result gives an affirmative answer.

\vspace{0.2cm} 

\begin{prop} \label{prop:ValueFunc} 
Under Assumptions~\ref{assump:NMPC1}-\ref{assump:Estmu} and with $\sigma= L_u + L_1$ it holds for $(x,u) \in \Gamma \times \mathcal{U}$ :
\begin{align} \label{eqn:DecreaseValueFunc} 
V(f(x,u)) \leq V(x) - l(x,u) + \sigma \| u - u^*(x)\|
\end{align}
\end{prop}

\vspace{0.15cm}
\textit{Proof:} 
The proof is given in Appendix~\ref{appx:ProofProp1}.

\vspace{0.15cm}
The constant $L_u = L_2L_4(1+L_3)$ is a composition of local Lipschitz constants, where $L_2$ and $L_3$ arise from the OCP (\ref{dfn:OP}) and $L_4$ arises from the Lipschitz continuity of $f(x,u)$ over the set $\Gamma \times \mathcal{U}$. Following Proposition \ref{prop:ValueFunc}, the difference in the decrease from $V(x)$ to $V(f(x,u))$ when applying the suboptimal input $u$, is the stage cost $l(x,u)$ plus the extra term $\sigma\| u - u^*(x)\|$, which acts like an additive disturbance on the optimal value function decrease. We now establish a condition which guarantees that $\sigma\| u - u^*(x)\| < l(x,u)$.

With Theorem~\ref{thm:RealTimeIt} and the stopping criterion in  Algorithm~\ref{algo:Algorithm1} $StopCrit(x, \bm{\mu}) = \frac{\sqrt{1- \epsilon^2}}{\sigma} \, l(x,u)$, we can show after $i$ steps: 

\vspace{-0.3cm}

\begin{align*}
\|u-u^*(x)\| &= \|\mu^i_0-\mu_0(x)^*\|  \\
&\leq \|\bm{\mu}^i-\bm{\mu}^*\|\\
& \leq\frac{1}{\sqrt{1-\epsilon^2}} \| \bm{\mu}^i- \Pi_{\mathcal{U}^N}\left[ \bm{\mu}^i - \gamma \nabla_{\bm{\mu}} h(x,\bm{\mu})\right]\| \\
&< \frac{1}{\sigma} \, l(x,u)
\end{align*}
\vspace{-0.3cm}

Thus, it holds that $\sigma\|u-u^*(x)\| < l(x,u)$ and by (\ref{eqn:DecreaseValueFunc}), $V(f(x,u)) < V(x)$. Consequently, the OCP (\ref{dfn:OP}) is recursively feasible as $\forall x \in \Gamma$ it holds $f(x,u) \in\Gamma$. Moreover, due to the strict decrease the system is asymptotically stable which completes the proof of Theorem~\ref{thrm:ValueFuncDecrease}.

\vspace{0.15cm}

\section{REAL-TIME PROJECTED GRADIENT-BASED NMPC FOR ANESTHESIA} \label{sct:Anesthesia}

\re{The real-time NMPC approach presented in the previous section is specially well-suited for drug infusion procedures, like anesthesia control, since it allows to include strict input constraints on the drugs, and is computationally efficient, which enables it to perform fast feedback-based control loops. In this section section we will demonstrate the applicability of our method on an anesthesia control problem.} The goal of anesthesia is for a patient to remain unconscious, calm, with little pain and a good maintenance of homeostasis (a stable internal body environment) and hemodynamic stability (stable blood flow dynamics) throughout a surgery. To achieve these goals a hypnotic and an analgesic drug is administered during intravenous anesthesia. Hypnotics induce unconsciousness and analgesics decrease sensitivity to pain. Most commonly, propofol is used as an hypnotic and remifentanil as an analgesic agent. Although, intravenous anesthesia is a multi-input multi-output problem, the focus lies on multi-input single-output control of depth of hypnosis due to a lack of a reliable analgesic sensors~\cite{eskandari2020extended}. The prevailing measure of depth of hypnosis is the bispectral index (BIS) which is based on phase coupling of different frequencies in the electroencephalogram. A fully awake patient has an index of 100 whereas an index of 0 corresponds to no brain activity. General anesthesia is achieved for a BIS of 40 to 60. Different types of disturbances arise during anesthesia which are often linked to sudden pain, i.e., a larger incision performed by the surgeon, or periods of very little stimulation, i.e., planning of the next steps in the surgery~\cite{dumont2012closed}. These disturbances must be handled efficiently to avoid the BIS leaving the target range of 40 to~60.


\subsection{Modelling}
\begin{figure}[b]
        \centering
        \includegraphics[width=0.485\textwidth]{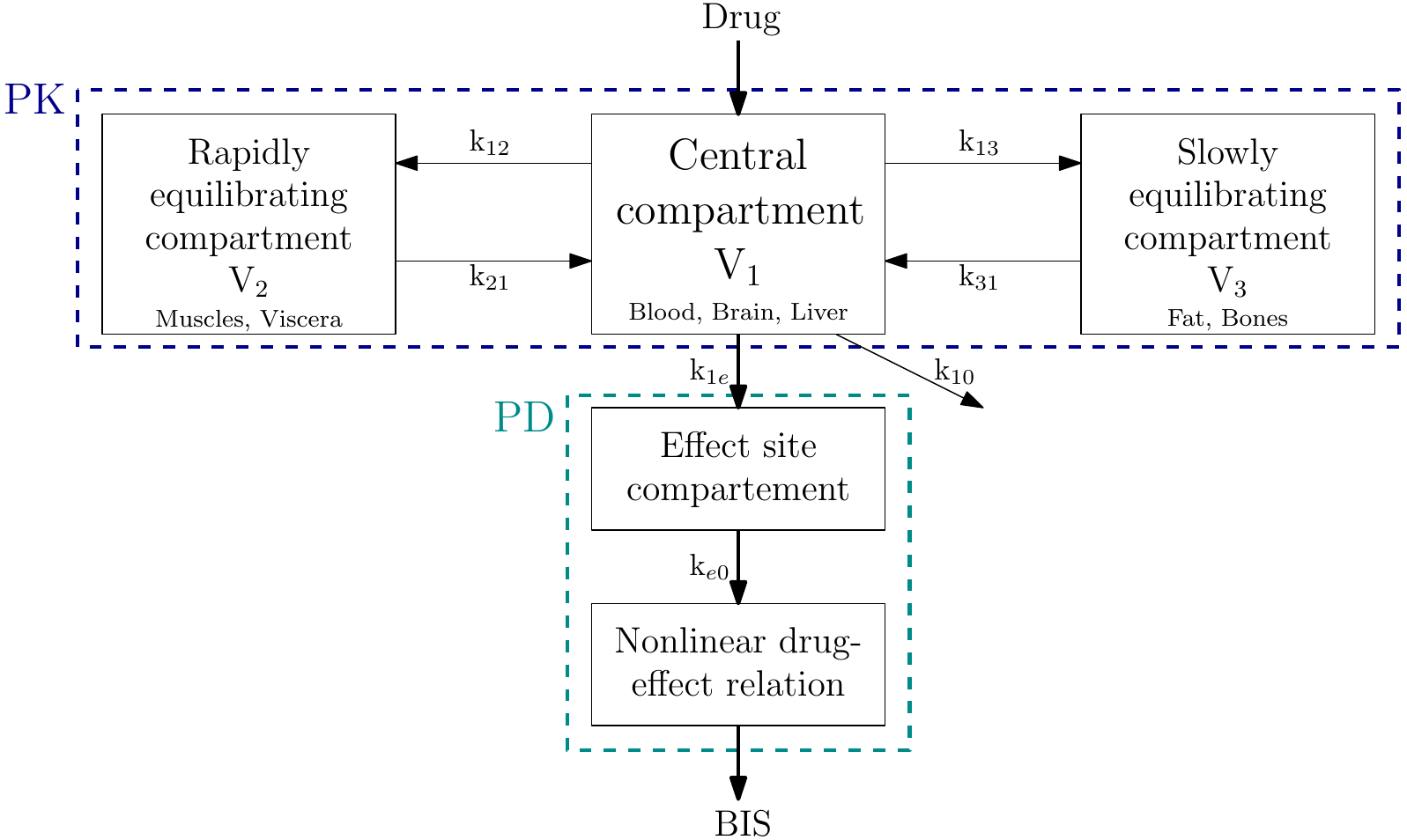} 
        \caption{Schematic of the linear PK compartment model and the nonlinear PD effect model, adapted from \cite{bibian2005introduction}.}
\label{fig:PKPDmodel}
\end{figure}

Modelling the distribution and effect of hypnotics and analgesics is usually done using pharmacokinetic (PK) and pharmacodynamic (PD) models. 
An overview of the PKPD model and the relationship from drug infusion (input $u$) to BIS effect (output $y$) is shown in Figure~\ref{fig:PKPDmodel}.

The linear PK model is based on a four-compartment model of the human body. It estimates drug concentration in each compartment and diffusion rates between compartments. For propofol infusion the Marsh~\cite{marsh1991pharmacokinetic} and Schnider~\cite{schnider1998influence} models are both commonly used in clinical practice. The Schnider model is not to be used with obese patients but incorporates age as an additional covariate which is advantageous~\cite{sahinovic2018clinical} and is therefore used here. The standard PK model for remifentanil is the Minto model~\cite{minto1997influence}.

The PK continuous-time state-space description for propofol and remifentanil is the following:

\vspace{-0.3cm}
{\begin{small}
\begin{equation*}
\begin{bmatrix}
\dot{C^p}\\
\dot{C^r}
\end{bmatrix}(t)=
\begin{bmatrix}
Ac^p & 0\\
0 & Ac^r 
\end{bmatrix}
\begin{bmatrix}
C^p\\
C^r
\end{bmatrix}(t) + 
\begin{bmatrix}
Bc^p & 0\\
0 & Bc^r
\end{bmatrix} 
\begin{bmatrix}
u^p\\
u^r
\end{bmatrix}(t),
\end{equation*}
\end{small}}

\noindent where $C^p = \begin{bmatrix}
C_1^p & C_2^p & C_3^p & C_e^p\\
\end{bmatrix}^T$ corresponds to the propofol concentration in the three different compartments and the effect site compartment. The same holds for $C^r$. The state-space matrices are composed of the transfer rates which are a function of patient specific variables, i.e, 

\vspace{-0.3cm}
{\begin{small}
\begin{align} \label{defn:TransRates}
Ac^p &= \begin{bmatrix}
-(k_{12}^p + k_{13}^p + k_{10}^p) & k_{21}^p & k_{31}^p & 0\\
k_{12}^p & -k_{21}^p & 0 & 0 \\
k_{13}^p & 0 & -k_{31}^p  & 0\\
k_{1e}^p & 0 & 0 & -k_{e0}^p
\end{bmatrix}\\ \nonumber
Bc^p &= \begin{bmatrix}
1 & 0 &  0 &  0 \\
\end{bmatrix}^T
\end{align}
\end{small}}
\vspace{-0.3cm}

\noindent \re{and similarly for $Ac^r$ and $Bc^r$ with the corresponding transfer rates for remifentanil $k^r$}. The nonlinear pharmacodynamics describe the interaction and the combined contribution of propofol and remifentanil to the observed effect BIS \cite{kern2004response}.

\vspace{-0.3cm}

{\begin{small}
\begin{align}\nonumber
&\text{BIS}(C_e^p, C_e^r) \\ 
&=  E_0 - E_{max} \, \frac{\left(\frac{C^p_{e}}{C^p_{50}} + \frac{C^r_{e}}{C^r_{50}} + \beta \times \frac{C^p_{e}}{C^p_{50}} \times \frac{C^r_{e}}{C^r_{50}}\right)^{\eta} }{\left(\frac{C^p_{e}}{C^p_{50}} + \frac{C^r_{e}}{C^r_{50}} + \beta \times \frac{C^p_{e}}{C^p_{50}} \times \frac{C^r_{e}}{C^r_{50}}\right)^{\eta} + 1}. \label{BISequ}
\end{align}
\end{small}}

The nominal model constants are taken from~\cite{kern2004response} and~\cite{sahinovic2018clinical}: 

\begin{itemize}
\item \re{($C_{50}^p$)} propofol drug concentration producing 50\% of the maximal effect:  1.8 $\mu$g/ml.
\item \re{($C_{50}^r$)} remifentanil drug concentration producing 50\% of the maximal effect:  12.5 $\mu$g/ml.
\re{\item ($E_0), (E_{max})$ baseline level and maximal effect: 100}.
\item ($\eta$) slope of the pharmacodynamic response curve: 3.76.
\item ($\beta$) interaction parameter for the two drugs: 5.1.
\end{itemize}

\subsection{Simulation Results}

\re{As we propose a novel control approach based on gradient-based real-time iterations which only approximates the MPC input, it is essential to demonstrate that nonetheless benchmark performance criteria for closed-loop anesthesia control given in~\cite{dumont2012closed} are met}:

\begin{enumerate}[label={\arabic*)}]
\item Rise time at induction of 3-4 min.
\item Overshoot less than 10-15\%.
\item During maintenance the BIS should stay within
10 points of the target in about 85\% of the time.
\item For an output disturbance (i.e., sudden analgesic arousal) the patient's response should be suppressed within 2 min without inducing oscillations.
\end{enumerate}

We test the closed-loop behaviour of these methods in two realistic clinical scenarios proposed in \cite{heusden2014safety} which are used as a benchmark for controller validation. We use a sampling time of 0.1 min and a NMPC prediction horizon of 25 steps. All transfer rates in (\ref{defn:TransRates}) can be computed from patient parameters such as weight, height and others. To estimate future concentrations a Kalman filter can be used which has proven to work well for anesthesia control~\cite{krieger2014model}. The infusion rates are both initialised with 1~$\frac{\text{mg}}{\text{min}}$ and 1~$\frac{\mu \text{g}}{\text{min}}$ for  propofol and remifentanil respectively. We adopt a quadratic stage-cost which regulates the input and penalizes deviations from the output reference $y_{ref}= BIS_{ref}= 50$,  {\re{$l(u,y) = 0.5 u^T R u + \rho/2 (y_{ref} -y)^2$}. The stepsize and cost function weights are found through tuning as $\gamma = 10^{-3}, \rho = 10$, and $R = \left[ \begin{smallmatrix} 1 & 0 \\ 0 & 1000 \end{smallmatrix}\right]$. The weight on the infusion rate of remifentanil is set much higher as it's primary use shall be analgesia rather than sedation. Exemplary patient profiles are taken from \cite{nogueira2019positive}. Time varying input constraints are required for anesthesia as drug infusion during induction (first 10 minutes) is much higher than during maintenance. The following parameters are used here \cite{jin2017coordinated}:

\begin{itemize}
\item Induction: $u_p  \leq 4\, \frac{\text{mg}}{\text{kg} \,\text{min}}$, $u_r \leq 0.36\,  \frac{\mu \text{g}}{\text{kg} \,\text{min}}$.
\vspace{0.1cm}
\item Maintenance: $u_p \leq 0.8 \frac{\text{mg}}{\text{kg} \,\text{min}}$,  $u_r \leq0.07\frac{\mu \text{g}}{\text{kg} \,\text{min}}$.
\end{itemize} 

The novel projected gradient-based controller successfully drives the patient to the BIS reference of 50, as seen in Figure~\ref{fig:Simulation2}~(a). The stopping criterion derived theoretically is overly conservative as the Hill-equation in (\ref{BISequ}) leads to a large Lipschitz constant $L_3$. Instead we can show in Figure~\ref{fig:Simulation2}~(a) that the controller is already stabilizing with only 10 iterations. While increasing the number of iterations the response becomes faster, and the steady-state error decreases. There is always a minimum natural physiological response time to drug infusions. The increase in rise time with only 10 iterations can be explained by the shape of the Hill-equation relating drug concentrations to BIS effect~\cite{kern2004response}. The gradients $\nabla_{C_e^p}\text{BIS}(C_e^p,C_e^r)$ and  $\nabla_{C_e^r}\text{BIS}(C_e^p,C_e^r)$ are nearly zero until a threshold concentration is reached. The controller with 50 iterations offers a good compromise between performance and computational complexity. The rise-time and overshoot criteria are fulfilled with significantly less iterations than the 1000 iterations of the ``close-to-optimal'' controller. 


To evaluate the disturbance rejection performance of the real-time NMPC scheme, two output disturbances are applied for one minute. First a positive disturbance increasing the BIS by ten points is applied which could be a sudden increase in pain. Then, a decrease in the BIS is simulated corresponding to a period of little stimulation. As shown in Figure~\ref{fig:Simulation2}~(b), in both cases the patient's response is suppressed  within two minutes, as specified by the performance criteria. \re{In addition, in Figure~\ref{fig:Simulation2}~(c) we demonstrate that the real-time NMPC scheme successfully drives the BIS to the reference value of 50 even under model uncertainties of 10\% to 30\%.}

\begin{figure}
         \includegraphics[width=0.495\textwidth]{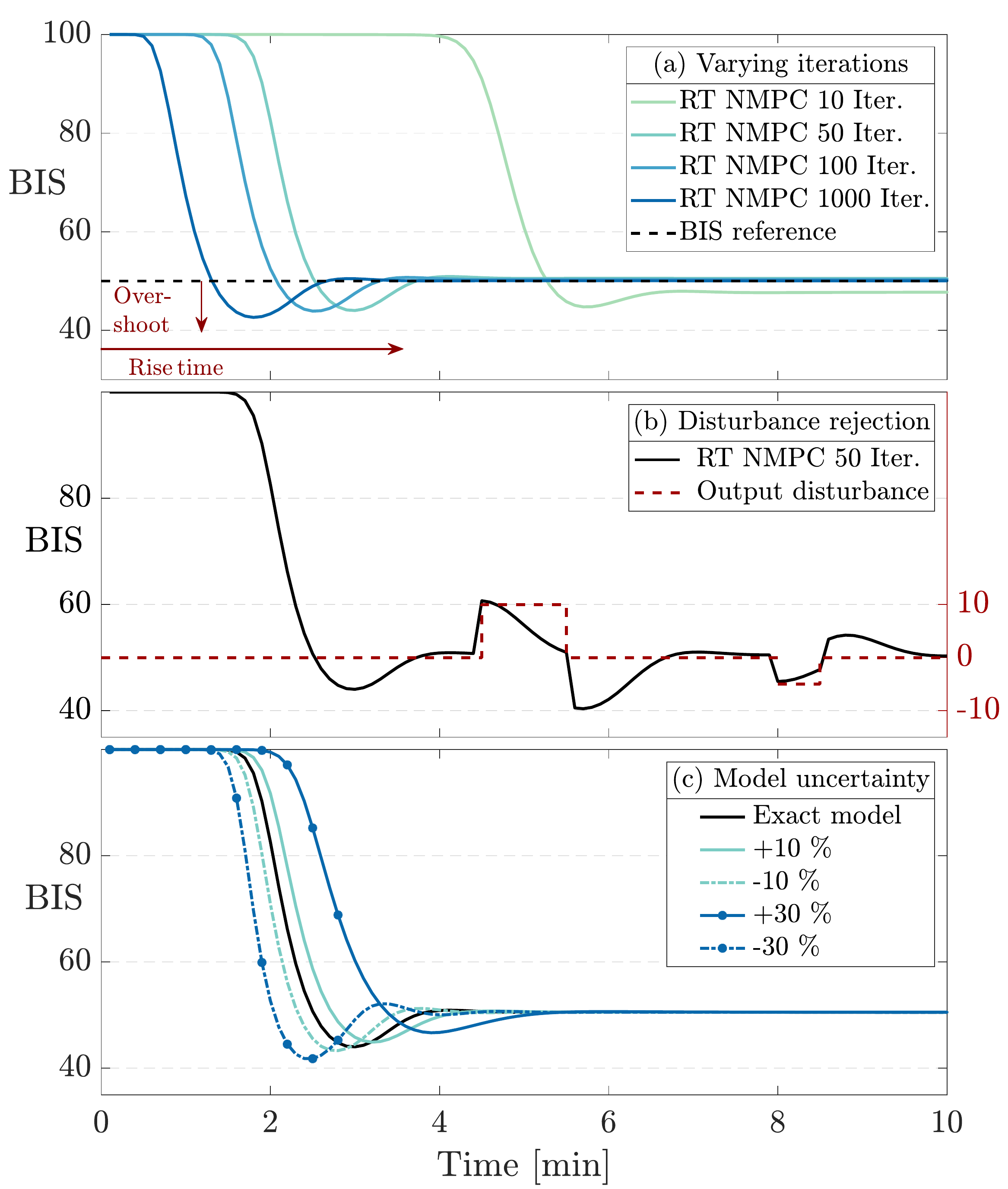} 
        \caption{(a) The BIS response curve with the real-time NMPC controller using varying number of iterations. (b) The disturbance rejection with 50 real-time NMPC iterations. (c) The control performance under model uncertainty (10\% to 30 \%) with 50 real-time NMPC iterations.}
\label{fig:Simulation2}
\end{figure}

\section{CONCLUSIONS}
\label{sct:Conclusions}

In this paper we have introduced a novel real-time iteration scheme for nonlinear model predictive control (NMPC) based on projected gradient descent. By taking steps along the gradient of the optimal NMPC cost, a complex optimization problem can be solved while fulfilling input constraints. Stability of the new scheme is proven by showing a decrease in the optimal value function at each iteration. This is achieved by bounding the difference between the optimal and suboptimal input such that the system does not diverge too much from the optimal trajectory.\\ The potential of the novel projected gradient-based NMPC scheme is demonstrated in an anesthesia control application. The controller fulfils commonly defined rise-time, overshoot and disturbance rejection criteria with only 50 iterations, The number of iterations can be reduced further by increasing the NMPC time horizon and improving initialisation.

\appendix

\subsection{Proof of Proposition  \ref{prop:ValueFunc}}
\label{appx:ProofProp1}

We begin by relating the change in the optimal value function $V(f(x,u))$ to the change of $u$.  
\vspace{-0.1cm}
\begin{align} \nonumber
&\| V(f(x,u))- V(f(x,u^*(x))) \| \\ \nonumber
&=\| h(f(x,u),\bm{\mu}^*(f(x,u)))  \\\nonumber 
& -h(f(x,u^*(x)), \bm{\mu}^*(f(x,u^*(x))))\|\\ \nonumber
&\leq  L_2\|f(x,u) - f(x,u^*(x))\| \\ 
& \quad + L_2 \|\bm{\mu}^*(f(x,u))- \bm{\mu}^*(f(x,u^*(x)))\| \label{eqn:DerivationL1}
\end{align}

Next, we investigate the sensitivity of the optimizer value $\bm{\mu}^*(x)$ to the parameter $f(x,u)$. Following \cite[Theorem 2.3.3]{jittorntrum1978sequential} and with Assumption~\ref{assump:Estmu} fulfilled, the optimizer value $\bm{\mu}^*(x)$ is $L_3$ Lipschitz continuous $\forall f(x,u)$ in the set $\Gamma \times \mathcal{U}$:  
\vspace{-0.8cm}

\begin{align} \nonumber 
&\|\bm{\mu}^*(f(x,u))- \bm{\mu}^*(f(x,u^*(x))) \| &\\
&\quad \leq  L_3 \| f(x,u) - f(x,u^*(x)) \|.
\label{eqn:DerivationL2}
\end{align} 
\vspace{-0.5cm}

Finally, the dynamics $f(x,u)$ are assumed to be continuously differentiable and thus they are $L_4$ Lipschitz on the compact sets $\Gamma$ and $\mathcal{U}$. Combining (\ref{eqn:DerivationL1}) and (\ref{eqn:DerivationL2}) gives:
\vspace{-0.3cm}

\begin{align}\nonumber
& \|V(f(x,u))- V(f(x,u^*(x))) \| \\ \nonumber
& \leq L_2 (1 +  L_3)\| f(x,u) - f(x,u^*(x)) \|   \\\nonumber
& \leq \underbrace{L_2L_4(1+L_3)}_{L_u} \| u - u^*(x) \|.
\label{eqn:DerivationMu}
\end{align}
\vspace{-0.3cm}

The following derivation connects all previous parts to arrive at the exact expression in Proposition \ref{prop:ValueFunc}. 
\vspace{-0.05cm}
\begingroup
\addtolength{\jot}{0.5em}
\begin{align*}
& V(f(x,u)) + \underbrace{V(f(x,u^*(x))) - V(f(x,u^*(x)))}_{=0}\\
& \leq \underbrace{V(f(x,u^*))}_{\leq V(x)-l(x,u^*(x))}+ \underbrace{\| V(f(x,u))-V(f(x,u^*(x)))\|}_{\leq L_u \| u-u^*(x)\|}\\
& \leq V(x) - l(x,u^*(x))+ L_u \| u-u^*(x)\|\\
& \leq V(x) - l(x,u) + L_u \| u-u^*(x)\| + \underbrace{\|l(x,u) -l(x,u^*(x))\|}_{\leq L_1 \|u-u^*(x)\|}\\
& \leq V(x) -l(x,u) +(L_u+L_1) \| u-u^*(x)\|\\
& = V(x) -l(x,u) + \sigma\| u-u^*(x)\|
\end{align*}
\endgroup

Thus, it holds that $V(f(x,u))\leq V(x) -l(x,u) + \sigma(\| u-u^*(x)\|)$. Here we made use of the $L_1$ Lipschitz continuity of the running cost $l(x,u)$ for $u \in \mathcal{U}$ given by Assumption~\ref{assump:RunningCost}.

\vspace{0.15cm}


\printbibliography

\end{document}